\documentclass{ws-mpla}

\newcommand{\Mpl}{M_\mathrm{Pl}}
\newcommand{\Ve}{V_\mathrm{eff}}

\newcommand{\bmat}{\beta_\mathrm{m}}
\newcommand{\bgam}{\beta_\gamma}
\newcommand{\rhom}{\rho_\mathrm{m}}
\newcommand{\meff}{m_\mathrm{eff}}
\newcommand{\Pgc}{{\mathcal P_\mathrm{pr}}}
\newcommand{\tp}{\tau_\mathrm{pr}}

\newcommand{\fesc}{f_\mathrm{esc}} 
\newcommand{\fvol}{f_\mathrm{vol}}
\newcommand{\edet}{\epsilon_\mathrm{det}} 

\newcommand{\fgam}{F_\gamma}
\newcommand{\faft}{F_\mathrm{aft}}
\newcommand{\ltot}{\ell_\mathrm{tot}}
\newcommand{\Gd}{\Gamma_\mathrm{dec}}
\newcommand{\Agam}{{\vec \Psi}_\gamma}
\newcommand{\Aphi}{{\Psi}_\phi}

\begin{document}

\markboth{J. H. Steffen \& A. Upadhye}
{GammeV Experimental Suite}

\catchline{}{}{}{}{}


\title{THE GAMMEV SUITE OF EXPERIMENTAL SEARCHES FOR AXION-LIKE PARTICLES}

\author{\footnotesize JASON H. STEFFEN}

\address{Fermilab Center for Particle Astrophysics\\
P.O. Box 500, MS 127\\
Batavia, IL 60506, USA
}

\author{AMOL UPADHYE}
\address{Kavli Institute for Cosmological Physics, Enrico Fermi Institute\\
University of Chicago\\
Chicago, IL 60637, USA
}

\maketitle

\pub{Received (Day Month Year)}{Revised (Day Month Year)}

\begin{abstract}
We report on the design and results of the GammeV search for axion-like particles and for chameleon particles.  We also discuss plans for an improved experiment to search for chameleon particles, one which is sensitive to both cosmological and power-law chameleon models.  Plans for an improved axion-like particle search using coupled resonant cavities are also presented.  This experiment will be more sensitive to axion-like particles than stellar astrophysical models or current helioscope experiments.

\keywords{Axion; Chameleon; Paraphoton; Laser Experiments}
\end{abstract}

\ccode{PACS Nos.: 12.20.Fv, 14.70.Bh, 14.80.Mz, 95.36.+x}

\section{Axion-like Particles and Laser Experiments}

Originally put forward as a possible explanation for the apparent lack of CP violation in strong force interactions,~\cite{peccei,wilczek,weinberg} the axion and axion-like particles (ALP) may be ubiquitous components of the Universe.  A massive axion remains a viable, and well motivated dark matter candidate,~\cite{abbott1983,preskill1983,dine1983} string theories often contain ALP fields,~\cite{witten1985,kain2006} and scalar fields are common suspects in theories of dark energy.~\cite{ratra1988,peebles1988} (In this work we broadly define an ALP as either a scalar or a pseudoscalar particle but reserve the term \textit{axion} specifically to refer to the axion of QCD.)

The most common experimental approach to detecting the presence of ALP's is via the coupling of an ALP to two photons.  For ALP's with keV energies, such as would be produced in the center of the Sun, this coupling to photons would cause Bragg diffraction of the ALP off of the atomic nuclei in a crystal lattice.~\cite{yoo2009,avigone,morales,bernabei}  Galactic ALP dark matter particles could drive electromagnetic (typically radio frequency) cavities as the particles stream through.~\cite{admx,rydberg}  Or, given an ambient electric or magnetic field, an ALP will undergo oscillations with a photon state and develop a nonzero photon amplitude.

Two means of searching for ALP-photon oscillations are, first, to convert ALPs produced in the Sun into photons with a helioscope, as is done with CAST~\cite{Andriamonje:2007ew} and Tokyo~\cite{minowa2008}, and second, to use a pair of magnets separated by an optical barrier, such that a photon from the a source can be seen through the barrier by oscillating into an axion which reconverts into a photon on the other side.  This second approach, dubbed ``light-shining-through-walls'' or LSW,~\cite{VanBibber:1987rq} has the advantage that the results do not rely on any model of either the Sun or the galaxy (for solar or dark matter ALPs respectively)---but it suffers from the severe limitation of needing both to make ALPs and then to regenerate photons, each of these conversions occuring through an inherently weak interaction.  Nevertheless, such LSW experiments are important and can be sensitive to other types of particles.~\cite{Jaeckel:2006xm,ahlers2008}  Furthermore, we shall see some ways that the sensitivity to ALPs can be enhanced.

The original LSW experiment was conducted by a collaboration between Brookhaven National Laboratory, Fermilab, the University of Rochester, and the University of Trieste (BFRT).~\cite{Cameron:1993mr}  They used two 4.4 meter, 3.25T dipole magnets and used an optical delay line to increase the number of passes of the laser beam on the production side of the experiment.  While they saw no evidence for new particles there were regions where, due to the fixed length of the magnets, they were not sensitive to ALPs.  This insensitivity occurs when the length of the magnet corresponds to an integer multiple of half the ALP-photon oscillation length---the probability of oscillating into an axion vanishes.

Roughly a decade after the BFRT experimental results were published, there was renewed interest in conducting this type of experiment.  In 2006, the PVLAS experiment reported a signal in a photon oscillation experiment [\refcite{Zavattini:2005tm}] that was consistent with the presence of a new spin-0 particle [\refcite{Raffelt:1987im}].  While they were not able to identify the parity of the particle, their measurements gave a consistent picture of a low-mass particle $m_\phi\sim 1.2\mbox{ meV}$ with a fairly large, two-photon coupling near $g \sim 2.5\times 10^{-6} \mbox{ GeV}^{-1}$.  Nominally such a particle would have been seen by the BFRT experiment.  However, the parameters inferred from PVLAS happened to coincide with a region of insensitivity of BFRT.  Shortly after the announcement of PVLAS, several LSW experiments, including the GammeV experiment,~\cite{chou2008} set out to test the particle interpretation of the PVLAS signal.~\cite{Patras}

While designing the GammeV experiment, we realized (as did two independent groups [\refcite{Ahlers:2007st,Gies:2007su}]) that with minor modifications to the apparatus the experiment could probe for so-called chameleon particles as well.  Chameleon particles are scalars or pseudoscalars with a matter coupling and a nonlinear self interaction.  These characteristics give the field an environment-dependent effective mass.  This environmental dependence allows the chameleon particles to have a small mass in a vacuum chamber, but large masses in generic terrestrial experiments.  Such particles could satisfy the particle interpretation of PVLAS while evading constraints on generic scalar particles that arise from torsion pendulum experiments;~\cite{Adelberger2007} they may also explain some of the observed light polarization effects seen in galactic and extragalactic sources.~\cite{burrage2009}

One strong motivation to search for chameleon particles is that their low mass in intergalactic space allows chameleons to play an important role cosmologically [\refcite{chamKW1,chamKW2}]---they may be the source of the observed cosmic acceleration.  The afterglow signature is remarkable in that it allows an experimentalist to probe non-cosmological consequences of dark energy in a laboratory setting.  Scalar field models dominate dark energy theory, and yet it is difficult to distinguish among them using cosmological observations.  By searching for afterglow, the chameleon seach by GammeV is able to probe a region of dark energy parameter space that is complementary to constraints that arise from cosmological observations.

A large range of chameleon models, including the chameleon dark energy and power law chameleons considered here, can be described approximately by a power law relation between the effective mass and the ambient density, $m_{\text{eff}} \propto \rho^{\eta}$, where $\eta$ is typically of order unity.
Chameleons may also couple to photons via terms such as $\phi F^{\mu\nu} F_{\mu\nu}$ and $\phi {\tilde F}^{\mu\nu}F_{\mu\nu}$, for scalars and pseudoscalars, respectively; here, $F_{\mu\nu}$ is the electromagnetic field strength tensor and ${\tilde F}_{\mu\nu}$ its dual.  This electromagnetic coupling, as with the axion, allows photons to oscillate into chameleons and back in the presence of an external magnetic field.  A significant difference is that, rather than penetrating the walls or windows of the vacuum chamber as would a generic ALP, the chameleon's mass increases sharply as it approaches the wall and the chameleon particle reflects when its effective mass equals its energy.
Since a chameleon with an energy less than the effective mass inside a material will be completely reflected by that material, the chameleon will become trapped inside a ``jar'' of that material.  This is true even for materials such as glass, which are transparent to photons.  A closed chamber may be populated with chameleons by streaming photons through the chamber, via glass windows, in the presence of a magnetic field.  These chameleons would regenerate photons after the photon source was turned off, producing an afterglow.

\section{The GammeV Experiments}

The GammeV ALP and chameleon searches were designed around a superconducting, 5 Tesla, 6 meter, Tevatron dipole magnet.  Many other components of the experiment, including the high power light source and the magnet operation infrastructure, relied on the resources available at Fermilab.  Below we describe some of the technical aspects of the GammeV experiment as well as its results.  We also discuss the results of the GammeV search for photon-coupled chameleon particles.

\subsection{GammeV Axion-Like Particle Search}

Two important features of our LSW ALP search are a pulsed laser and a moveable barrier.~\cite{chou2008}  Our frequency-doubled, Continuum Surelite I-20 Nd:YAG laser provided 5 ns, 160 mJ pulses of 532 nm light at a rate of 20 Hz.  The pulsed nature of the light source provided a means to reduce drastically the effects of detector noise by requiring the signal to be coincident with the arrival time of the laser pulse.  A half-wave plate allowed us to test both polarizations of light---and hence to search for both scalar and pseuodoscalar ALPs.

The optical barrier (or ``wall'') for this experiment was a stainless steel cap welded onto a hollow cylinder.  This ``plunger'' was inserted into the warm bore of the magnet; the cap on the plunger partitioned the magnet into two regions with a combined length of 6 meters.  The advantage of the plunger design is that sliding the plunger to different locations changes the lengths of the two magnetic field regions---thereby changing the sensitivity of the experiment to ALP mass.  Using two positions of the plunger, the GammeV experiment was sensitive to all values of the ALP mass in the PVLAS region of interest.  Moreover, should a signal have been seen, the mass of the particle could be identified by making incremental adjustments to the position of the plunger and monitoring the relative strength of the signal.  A mirror mounted on the end of the plunger reflected incident laser light out of the magnet.  This reduced the energy deposited in the bore of the magnet, and allowed us to monitor in real time the laser power passing through the system.

Photons entering the magnetic field region oscillate into a mixture of photon and ALP.  At the mirror, the particles either reflect as pure photons or pass through as pure ALPs.  The ALPs then oscillate back into photons through the remaining magnetic field region inside the inner diameter of the hollow plunger.  Upon exiting the magnetic field region, the interaction ceases and the photon-ALP wavefunction is frozen until it strikes the Hamamatsu H7422P-40 photomultiplier tube (PMT).  The PMT detects single photons, and those which are coincident with the laser pulses are candidate signal events.  A high signal to noise ratio is achieved due to the short pulses emitted by the laser.

The photon-ALP transition probability may be written in convenient units as:
\begin{eqnarray}
\label{E:conversionprob1}
P_{\gamma \rightarrow \phi} &=& \frac{4 B^2\omega^2}{M^2 (\Delta m^2)^2} \sin^2 \left( \frac{\Delta m^2 L}{4\omega} \right) \\
\label{E:conversionprob2}
&\approx&  \frac{4 B^2\omega^2}{M^2 m_\phi^4} \sin^2 \left( \frac{m_\phi^2 L}{4\omega} \right) \\
 \label{E:conversionprob3}
&=& 1.5\times 10^{-11} \frac{(B/\mbox{Tesla})^2(\omega/\mbox{eV})^2}{(M/10^5 \mbox{ GeV})^2 (m_\phi/10^{-3} \mbox{ eV})^4} \nonumber \\
& & \times \sin^2 \left( 1.267 \frac{(m_\phi/10^{-3} \mbox{ eV})^2 (L/\mbox{m})}{(\omega/\mbox{eV})} \right)
\end{eqnarray}
where $B$ is the strength of the external magnetic field, $\omega$ is the inital photon energy, and $L$ is the magnetic oscillation baseline.   The mass-squared difference between the scalar mass and the effective photon mass, $\Delta m^2=m_\phi^2-m_\gamma^2$, characterizes the mismatch of the phase velocities of the photon wave and the ALP wave.  This phase mismatch determines the oscillation length.  While a photon does not truly gain mass in a normal dielectric medium, the phase advance can be treated as an effective imaginary mass 
$m_\gamma^2 = -2 \omega^2 (n-1)$,
where $n$ is the index of refraction.~\cite{vanBibber:1988ge}  The interior of both the warm bore and the plunger are pumped to vacuum pressures less than $10^{-4} \mbox{ Torr}$---giving a conservative estimate of $\sqrt{-m_\gamma^2} <  10^{-4} \mbox{ eV}$.  Therefore, the photon mass is assumed to be negligible starting with Eqn.~\ref{E:conversionprob2} for values of $m_\phi$ near the PVLAS region.  For values of $m_\phi$ smaller than about $10^{-5} \mbox{eV}$, the ALP mass is negligible and the constraints that we found are equally valid.

Equation~\ref{E:conversionprob3} shows that the meter-scale baseline provided by typical accelerator magnets is well-suited for probing the milli-eV range of ALP masses.  That equation also shows that a monochromatic laser beam and a fixed magnetic field length gives regions where a particular configuration is insensitive to some masses of ALP.  As stated, this is why the BFRT experiment would not have seen the PVLAS signal, even though they had similar sensitivity.  GammeV's plunger design allows us to change the oscillation baseline to effectively cover all values of the ALP mass.  For GammeV, the total probability of converting photons to ALPs and back varies as $\sin^2 \left(\frac {m_\phi^2 L_1}{4\omega} \right) \sin^2 \left(\frac{m_\phi^2 L_2}{4\omega} \right)$ where $L_1+L_2=6\mbox{ m}$.

To detect regenerated photons we use a $51\mbox{ mm}$ diameter lens to focus the beam onto the $5 \mbox{ mm}$ diameter GaAsP photocathode of the PMT.  The alignment of the optical system with the PMT is performed using a low power green helium-neon laser and a mock target.  We verified the alignment both before and after each major data-taking period by replacing the plunger with an open ended tube, re-establishing the vacuum, and firing the high power laser onto a flash paper target.  We measured an optical transport efficiency of $92\%$.  The factory-measured quantum efficiency of the photocathode is $38.7\%$ and the collection efficiency of the metal package PMT is estimated to be $70\%$.  The PMT pulses are amplified by 46 dB and then sent into a NIM discriminator.  We used a highly attenuated LED flasher as a single photon source in order to optimize the discriminator threshold to give 99.4$\%$ efficiency for triggering on single photo-electron pulses while also rejecting the lower amplitude noise.  The deadtime fraction due to possible multiple rapid PMT pulses is found to be negligible (0.001$\%$).  Thus, the total photon transport and counting efficiency is estimated to be $(25\pm 3)\%$.  Using this threshold, and the built-in cooler to cool the photocathode to $0^\circ$C, we measure a typical dark count rate near 130 Hz.

The coincidence counting is accomplished using two Quarknet boards [\refcite{quarknet,quarknetgps}] with 1.25 ns timing precision---both referenced to a GPS clock.  The clocks on the laser board and on the PMT board are synchronized using an external trigger from a signal generator.  We identified the time interval between the laser pulses and the PMT response by removing the plunger with the mirror, and allowing the laser to shine on the PMT through several attenuation stages: 1) two partially reflective mirrors, 2) a pinhole, and 3) multiple absorptive filters mounted directly on the aperture of the PMT module.  These attenuation stages reduce the $10^{18}$ photons per pulse emitted by the laser to approximately 0.005 photons per pulse arriving at the PMT (a 21 order-of-magnitude reduction that results in an approximately 0.1 Hz signal).  This signal is used both to calibrate the timing between the two Quarknet boards and to provide an \textit{in situ} test of the data acquisition system.

\begin{figure}[ht]
\centerline{\psfig{file=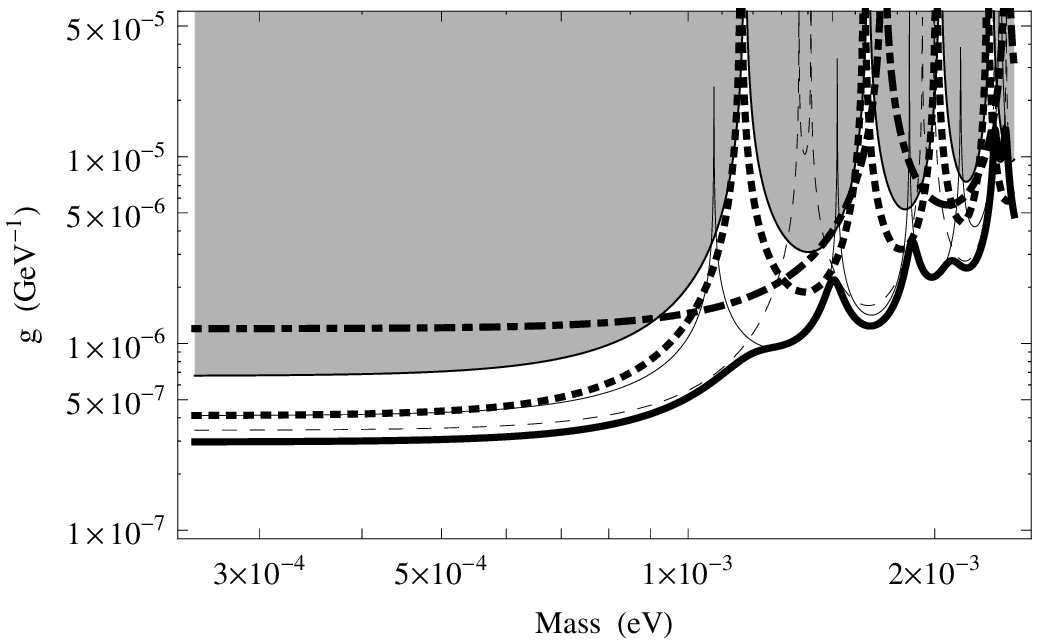,width=4in}}
\vspace*{8pt}
\caption{Constraints on scalar axion-like particle couplings to photons as a function of particle mass.  Constraints from the GammeV experiments are the thick solid curve, which uses data from both the center (thin dashed) and edge (thin solid) positions of the plunger.  The BFRT limits are shown as the shaded region.  The ALPS limits are the dotted curve.  The LIPPS experimental limits are dot-dashed (scalar only).\protect\label{gamscalar}}
\end{figure}

\begin{figure}[ht]
\centerline{\psfig{file=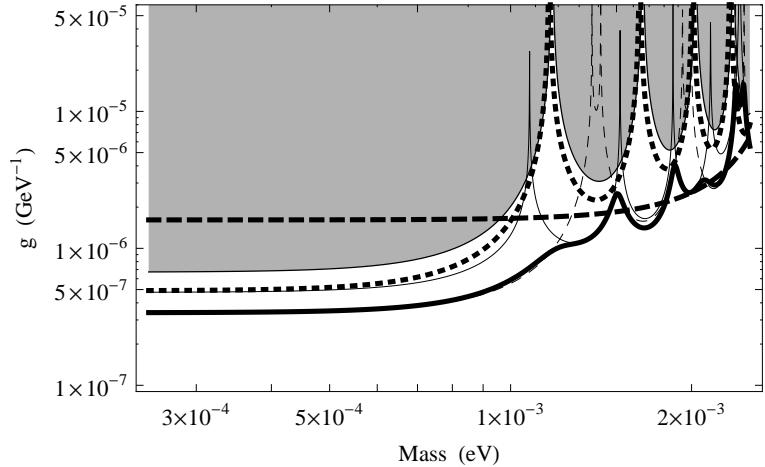,width=4in}}
\vspace*{8pt}
\caption{Constraints on pseudoscalar axion-like particle couplings to photons as a function of particle mass.  Constraints from the GammeV experiments are the thick solid curve, which uses data from both the center (thin dashed) and edge (thin solid) positions of the plunger.  The BFRT limits are shown as the shaded region.  The ALPS limits are the dotted curve.  The BMV limits are dashed (pseudoscalar only).\protect\label{gampseudo}}
\end{figure}

Since the low-mass ALPs are highly relativistic, the regenerated photons should arrive at the same time as the straight-through photons.  The coincidence window for our counting purposes was chosen to include 99\% of the time distribution of the laser pulses---giving a 10ns window.  The coincident background rate from the PMT is approximately $R_{\rm{noise}}=20\mbox{ Hz}\times 130\mbox{ Hz}\times 10\mbox{ ns} = 2.6\times 10^{-5} \mbox{ Hz}$.  As our data span roughly 5 hours, this gives an estimate of about one background photon.  Ultimately we measured the background rate using non-coincident data that surrounded the arrival of laser pulses and found results consistent with these expectations.  We took data in four configurations of our apparatus---two different plunger positions for each of the two laser polarizations.  The results of these four configurations are summarized in Table \ref{axiondata}.  Using the Rolke-Lopez [\refcite{Rolke:2004mj}] method to determine the exclusion limits for these data, we obtain the results shown in Figures \ref{gamscalar} and \ref{gampseudo}.
\begin{table}
\begin{center}
\label{axiondata}
\caption{Summary of data in each of the 4 configurations.}
\begin{tabular}{|l|c|c|c|c|}
\hline
Configuration&$\#$ photons&Est.Bkgd&Candidates&g[GeV$^{-1}$]\\
\hline
Horiz.,center&$6.3\times 10^{23}$&$1.6$&1&$3.4\times 10^{-7}$\\
Horiz.,edge&$6.4\times 10^{23}$&$1.7$&0&$4.0\times 10^{-7}$\\
Vert.,center&$6.6\times 10^{23}$&$1.6$&1&$3.3\times 10^{-7}$\\
Vert.,edge&$7.1\times 10^{23}$&$1.5$&2&$4.8\times 10^{-7}$\\
\hline
\end{tabular}
\end{center}
\end{table}

\subsection{Other Searches for Axion-Like Particles}

A number of other experimental probes for ALPs followed the announcement of the PVLAS results.~\cite{Patras}  Most of these efforts, like GammeV, were centered at national research facilities where suitable, and otherwise expensive, components were available.  Notable efforts include the BMV experiment [\refcite{Robilliard:2007bq}], LIPPS [\refcite{Afanasev2008}], OSQAR [\refcite{Pugnat:2007nu}], and ALPS [\refcite{alpsresults}].  The PVLAS collaboration continued their own study by rebuilding their apparatus.  While GammeV and the aforementioned experimental efforts were under way the PVLAS collaboration noted that they could not reproduce their results using the rebuilt instrument.~\cite{Zavattini:2007ee}  Nevertheless, the other efforts went forward and their results (also shown in Figures \ref{gamscalar} and \ref{gampseudo}) have since been reported.  We note that the OSQAR experimental limits are not shown in these figures.  This is because their limits rely on different assumptions for the propogation of low-energy ($\sim$eV) photons through a gas (see [\refcite{Pugnat:2007nu}] and [\refcite{leonhardt}] for a discussion).  Nevertheless, given their assumption they obtain similar sensitivity to the other ALP experiments.~\cite{Pugnat:2007nu,alpsresults}

\subsection{The GammeV Chameleon Particle Search}

With minor modifications to the ALP apparatus, the GammeV experiment searched for photon-coupled chameleon particles through the afterglow effect.  The modification to the apparatus was simply to remove the plunger that was used in the ALP seach, and to replace it with the hollow tube that we had used to align the laser.  We also inserted a mirror in front of the PMT to reflect the light back through the chamber.  This mirror accomplished two goals, it allowed us to keep the power detection system in place in the laser box and it doubled the number of photons passing through the magnetic field region---and hence the chameleon yield.

\begin{figure}[tb]
\begin{center}
\includegraphics[width=2.75in,angle=270]{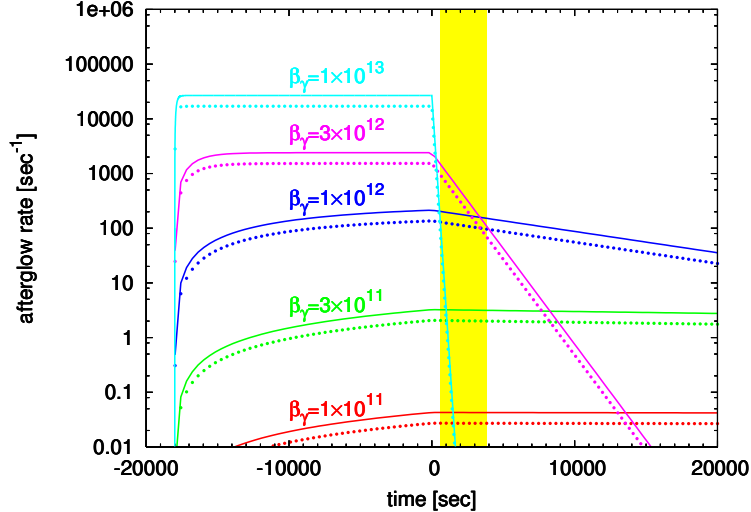}
\caption{Photon flux vs. time, for different chameleon-photon coupling strengths $\beta_\gamma \equiv M_\mathrm{Pl}/M$ and effective chameleon masses inside the chamber.  Solid and dotted lines correspond to $\meff=10^{-4}$~eV and $\meff = 5\times 10^{-4}$~eV, respectively.  The laser is turned off at time $t=0$; the shaded region represents our observation window.\label{fluxgraphic}}
\end{center}
\end{figure}

To populate the chameleon jar, the laser operated continuously for $\tp\approx 5\mbox{ h}$.  The 1 cm$^{-1}$ linewidth of the laser is sufficiently large to span the discrete energy levels of the trapped chameleons.  Following the filling stage the laser was turned off, and the PMT was uncovered after a brief (few minute) pause needed to prepare it for observation.  Data were taken in two separate runs, with the laser polarization either parallel or perpendicular to the magnetic field.  Each data gathering session was approximately one hour in length.  Figure~\ref{fluxgraphic} shows the expected afterglow rate as a function of time, during the filling and afterglow stages; our observation window for pseudoscalar particles is the shaded region.


A chameleon scalar field $\phi$ coupled to matter and photons has an action~\cite{chamcos}
\begin{equation}
S = \int d^4x {\Big (}-\frac{1}{2}\partial_\mu\phi\partial^\mu\phi - V(\phi) - \frac{e^{\phi/M_\gamma}}{4}F^{\mu\nu}F_{\mu\nu} + {\mathcal L}_\mathrm{m}(e^{2\phi/M_\mathrm{m}}g_{\mu\nu},\psi^i_\mathrm{m}) {\Big )}
\label{e:action}
\end{equation}
where $g_{\mu\nu}$ is the metric, $V(\phi)$ is the chameleon potential, and ${\mathcal L}_\mathrm{m}$ is the Lagrangian for matter.  We consider a universal coupling to matter $\bmat = \Mpl / M_\mathrm{m}$, and a photon coupling $\bgam = \Mpl / M_\gamma$, where $\Mpl = 2.4 \times 10^{18}$~GeV is the reduced Planck mass and $M_\mathrm{m}$ and $M_\gamma$ are the mass scales associated with the coupling descending from the theory for matter and photons respectively.  The non-trivial couplings to matter and the electromagnetic field induce an effective potential
\begin{equation}
\Ve(\phi,\vec x) = V(\phi) + e^{\bmat\phi/\Mpl} \rhom(\vec x) + e^{\beta_\gamma\phi/\Mpl} \rho_\gamma(\vec x),
\end{equation}
where we have defined the effective electromagnetic field density $\rho_\gamma = \frac{1}{2}(|\vec B^2|-|\vec E|^2)$ (for scalars) or $\rho_\gamma = \vec E \cdot \vec B$ (for pseudoscalars) rather than the energy density.  The expectation value $\left< \phi \right>$, the minimum of $\Ve$ and thus the effective mass of the chameleon ($\meff \equiv \sqrt{d^2\Ve/d\phi^2}$ ), depends on the density of both background matter and electromagnetic fields.

Our experiment is sensitive to chameleons that satisfy two conditions, production and containment. Production requires chameleons to be light enough inside our chamber to allow coherent photon-chameleon oscillation, $\meff(\mathrm{chamber}) \ll \sqrt{4\pi \omega/L} = 9.8\times 10^{-4}$~eV.  Containment requires that chameleons are massive enough to reflect from the walls of the chamber.  Clasically, this implies $\meff(\mathrm{wall}) > \omega$; a semiclassical calculation shows that $\meff(\mathrm{wall})-\omega > 10^{-6}\textrm{ eV} \ll \omega$ is sufficient to prevent significant losses through tunneling as the chameleon reflects $\sim 10^{13}$ times during our data runs.  In GammeV, the weakest (that is, least dense) part of the wall is our pumping system, which consists of a turbomolecular pump connected to a roughing pump.  Because the roughing pump exhausts to the room, chameleons must reflect from the air at the intake of the roughing pump, with pressure $P_\mathrm{pump} = 1.9\times 10^{-3}$~Torr, in order to remain in the chamber.  The turbo pump merely acts as extra volume, $V = 0.026 \mathrm{ m}^3$, to the highly relativistic chameleons.

If the relation between the chameleon mass and the local density is a power law, which we parameterize as $\meff(\rho) = m_0 (P(\rho)/P_\mathrm{pump})^\eta$, then our production and containment conditions can be expressed concisely as $\omega < m_0 < \sqrt{4\pi\omega/L}(P_\mathrm{pump}/P_\mathrm{chamber})^\eta$.  Here, $P_\mathrm{chamber} \approx 10^{-7}$~Torr is the vacuum pressure maintained inside the chamber during the chameleon production and afterglow phases; we are therefore sensitive to chameleons with $\eta \gtrsim 0.8$.  This power law scaling neglects the dependence of $\meff$ on the magnetic field energy density; since in the GammeV apparatus $\rho_\mathrm{gas} \sim \frac{1}{2}B^2 \sim 10^{-13}\mathrm{ g/cm}^3$, $\meff$ varies the most--and our experiment is the most sensitive--when $\bmat \gg \bgam$.

In terms of the coupling $\bgam$, and $\meff$ in the chamber, the chameleon production probability [\refcite{Sikivie:1983ip,Sikivie:1985yu,Raffelt:1987im}] per photon is
\begin{equation}
\Pgc = \frac{4\bgam^2B^2\omega^2}{\Mpl^2 \meff^4}
\sin^2\left(\frac{\meff^2 L}{4\omega}\right).
\label{e:Pgc}
\end{equation}
A particle that has just reflected from one of the chamber windows is in a pure chameleon state.  As the chameleon particles bounce off of the imperfectly aligned windows and chamber walls, the chameleon momenta become isotropic.  Each time the chameleon passes through the magnetic field region, 
it has a small probability of oscillating into a photon and escaping.
In the small mixing angle limit, the photon amplitude $\Agam$ due to this oscillation is given by
\begin{equation}
\left(-\frac{\partial^2}{\partial t^2} - k^2\right) \Agam
= \frac{k\bgam B}{\Mpl} {\hat k}\times(\hat x \times \hat k) \Aphi,
\label{e:Agam}
\end{equation}
where $|\Aphi| \approx 1$ is the chameleon amplitude, $k\approx \omega$ is the momentum, and $\hat k$ and $\hat x$ are unit vectors in the direction of the particle momentum and the magnetic field, respectively.  The chameleon decay rate corresponding to a particular direction $(\theta,\varphi)$ is $(|\Agam(\theta,\varphi)|^2 + {\mathcal P}_\mathrm{abs}(\theta,\varphi)) / \Delta t(\theta)$ evaluated at the exit window, where $\theta$ is the direction with respect to the cylinder axis, ${\mathcal P}_\mathrm{abs}$ is the total probability of photon absorption in the chamber walls, and $\Delta t(\theta) = \ltot/\cos(\theta)$ is the time required to traverse the $\ltot\approx 12.3\mbox{ m}$ chamber.  We model a bounce from the chamber wall as a partial measurement in which the photon amplitude is attenuated by a factor of $f_\mathrm{ref}^{1/2}$, where $f_\mathrm{ref}$ is the reflectivity.  The mean decay rate $\Gd$ per chameleon is found by averaging over $\theta$ and $\varphi$.  Although the cylinder walls are not polished, we assume a low absorptivity $1-f_\mathrm{ref}=0.1$ in order to overpredict the coherent build-up of photon amplitude over multiple bounces.  This overprediction of the decay rate results in a more conservative limit on the coupling constant.  In the limit of low $\meff(\mathrm{chamber})$, we obtain $\Gd = 9.0\times 10^{-5} (\bgam/10^{12})^2 (B/(5\mathrm{ Tesla}))^2$~Hz.

While the laser is on, new chameleons are produced at the rate of $\fgam \Pgc$, where $\fgam$ is the input photon rate,  and decay at the rate of $N_\phi \Gd$.  After a time $\tp$ the laser is turned off, and the chamber contains $N_\phi^\mathrm{(max)} = \fgam \Pgc \Gd^{-1} (1-e^{-\Gd \tp})$ chameleon particles.  For our apparatus, this saturates at $3.6\times 10^{12}$ for $\bgam \gtrsim 10^{12}$ and small $\meff$.   The contribution to the afterglow photon rate from non-bouncing chameleon trajectories is
\begin{equation}
\faft(t) = \frac{ \edet \fvol \fesc \fgam \Pgc^2 c}{\ltot \Gd}\left(1-e^{-\Gd\tp}\right) e^{-\Gd t},
\label{E:f_aft}
\end{equation}
 for $t\geq 0$, where $t=0$ is the time at which the laser is turned off.  The detector efficiency $\edet$ contains the same factors as those measured for the ALP search: $0.92$ optical transport efficiency, $0.387$ quantum efficiency, and $0.7$ collection efficiency of the PMT.  Because chameleons in the turbo pump region do not regenerate photons, we consider only the chameleons in the cylindrical chamber, which represents a volume fraction $\fvol=0.40$ of the total population.

\begin{figure}[tb]
\begin{center}
\includegraphics[width=2.75in,angle=270]{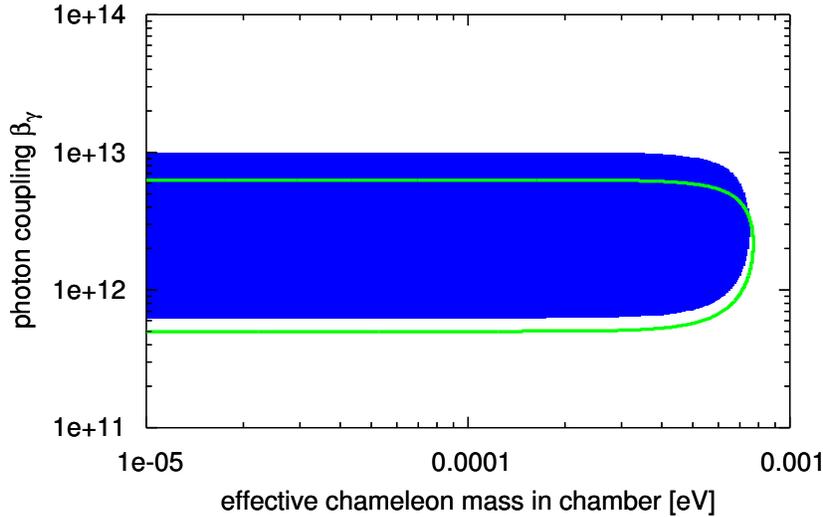}
\caption{Constraints ($3\sigma$) on the coupling of chameleon particles to photons as a function of the chameleon mass from the GammeV chameleon particle search.  The solid region is excluded for pseudoscalar chameleons, and the region inside the green curve for scalar chameleons. \label{chameleonresults}}
\end{center}
\end{figure}

In order to set conservative, model-independent limits, we consider only the afterglow from the fraction $\fesc=5.3\times 10^{-7}$ of chameleons which travel the entire distance $\ltot$ from entrance to exit windows without colliding with the chamber walls, and are focussed by the 51~mm lens onto the PMT.  While most of the photons that can reach the detector are on trajectories that bounce from the chamber walls, such collisions result in a model-dependent chameleon-photon phase shift which affects the coherence of the oscillation on bouncing trajectories.~\cite{Brax:2007hi}  Figure \ref{fluxgraphic} shows the prediction for the minimum afterglow signal---consisting of only the direct light---and attenuated by the fastest possible decay rate $\Gd$ in Eq.~\ref{E:f_aft}.  This afterglow rate is plotted for several values of the photon-chameleon coupling $\beta_\gamma$.  Non-observation of this underpredicted rate sets the most conservative limits.  The three-sigma constraints obtained from the chameleon data are shown in Figure \ref{chameleonresults}.

We also note that in more general models, chameleons could decay to undetectable daughter particles.  Furthermore, large chameleon self-interactions can lead to fragmentation and thermalization of chameleons, weakening our constraints.  We do not address those points here, but expect to at a later time (in conjunction with a second generation experiment described below).  Our goal in determining these results was to present them, as much as is feasible, in a manner that is independent of the chameleon model.

\subsection{Limits on Other Particles from GammeV}

In addition to ALPs and chameleons, laser experiments are sensitive to other types of particles such as paraphotons that couple to hidden sector particles and mini-charged particles that, due to couplings with dark sector particles, have electric charges much less than that of the electron.~\cite{ahlers2008}  For the case of photon-paraphoton couplings an ambient magnetic field is not needed to induce oscillations between the two particles.  Solar physics and cosmological constraints on these particles exist, yet both astrophysical probes are relatively less sensitive than laboratory probes for particle masses near one milli-eV.  Constraints on paraphotons and minicharged particles from GammeV and other experiments can be found in [\refcite{ahlers2008}].

\section{Avenues for Future Experiments}

The lessons learned from the first set of GammeV particle searches provide important starting points for improved experiments.  Two such experiments are currently being developed---one for ALPs and the other for chameleon particles.  In these instances the experiments are more optimized for one or the other particle and do not share the same apparatus or design.  The experiment first to be pursued is an improved search for chameleon particles.  The second, a significantly enhanced experiment to search for ALPs, requires much more technical development and will proceed on a longer timescale.

\subsection{The Chameleon Afterglow Search (CHASE)}

As discussed above and in [\refcite{Chou2009}], the original GammeV chameleon search suffered from four different technical limitations in probing for chameleon particles:
\begin{enumerate}
\item Destructive interference in our $L=6$~m magnetic field length limited us to chameleons lighter than $\sqrt{4\pi\omega/L} \approx 10^{-3}$~eV.
\item Systematic uncertainties in our PMT dominated our uncertainties.  This limited our ability to detect low $\bgam$.
\item Uncovering our PMT required several minutes after the laser was turned off, so our sensitivity to high-$\bgam$ (rapidly decaying chameleons) was diminished.
\item Our vacuum system limited our ability to trap chameleons with $\eta \lesssim 0.8$.
\end{enumerate}
CHASE improves our sensitivity considerably by addressing each one of these limitations.  Glass windows will divide our magnetic field region into compartments of different lengths $\sim 1$~m, and $\sim 5$~m.  These two compartments remove regions of insensitivity, much like the moveable plunger did for the original ALP search.  They also provide some sensitivity to larger mass chameleons, up to a few meV.  This improvement is especially significant because chameleon masses at the dark energy scale, $\rho_\mathrm{de}^{1/4} = 2.3 \times 10^{-3}$~eV, were inaccessible to our previous experiment.

Improvements to our optical system will allow us to push to both lower and higher $\bgam$.  By modulating the PMT signal (using a mechanical shutter), we can better monitor the detector noise.  Since systematic uncertainty in the PMT dark rate was our dominant source of noise in [\refcite{Chou2009}], our sensitivity to low afterglow rates will improve by roughly an order of magnitude in CHASE.  Meanwhile, making a more rapid transition between filling the cavity and collecting data (shortening the dead time by a factor of more than 100) we will improve our sensitivity to larger $\bgam$ by three orders-of-magnitude.  Our high-$\bgam$ constraints will be further improved with runs having lower magnetic fields, near 0.2 Tesla.  The low magnetic field slows conversion of chameleon particles to photons, this maintains a detectable population of chameleons while the transition to data acquisition occurs.

\begin{figure}[ht]
\centerline{\psfig{file=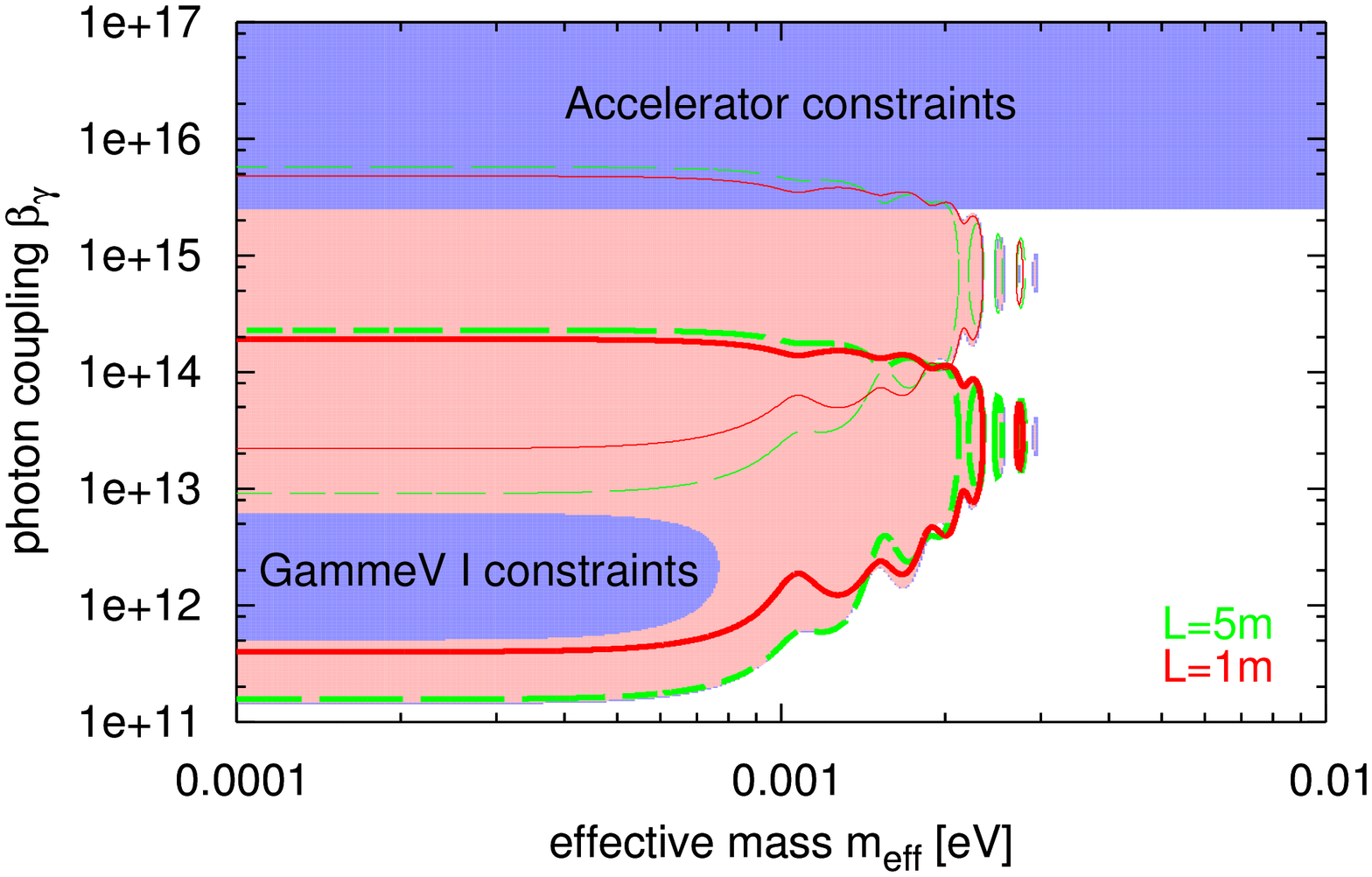,width=2.67in,angle=270}}
\vspace*{8pt}
\centerline{\psfig{file=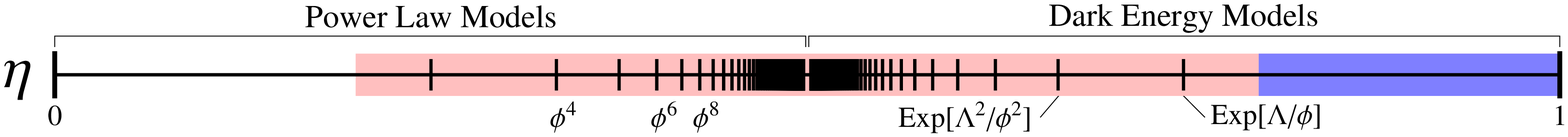,width=4in}}
\vspace*{8pt}
\caption{
Forecast CHASE constraints.  (top) Model-independent exclusion plot, for two different compartment lengths, 5 meters (dashed) and 1 meter (solid), and for two magnetic field strengths 5 Telsa (lower region) and 0.2 Tesla (upper region).  The combined constraints yield the shaded region.  (bottom) Range of potentials excluded. In both cases, current constraints are shown in blue/dark gray, and forecast CHASE constraints in pink/light gray.  Accelerator constraints are from [51].
\protect\label{chaselimits}
}
\end{figure}

Finally, improvements to our pumping system will take two forms.  First, our vacuum will be reduced by approximately three orders of magnitude from $\sim 10^{-7}$ torr to $\sim 10^{-10}$ torr.  Second, our vacuum system will not exhaust to the room as it did in the original experiment.  These improvements come through the use of ion pumps placed at strategic locations on the apparatus and cryogenic pumping within the magnetic field region (i.e. residual gasses will freeze to the bore of the magnet).  The fact that this system does not exhaust to the room will mean that chameleons need only bounce from the chamber walls, with densities $\rho \sim 1$~g/cm$^3$, rather than the much stronger condition that they bounce before the intake of our roughing pump, with $\rho \sim 10^{-9}$~g/cm$^3$.  Figure \ref{chaselimits} shows the anticipated sensitivity of CHASE to $\bgam$ and $m_\phi$ as well as the increased range in potentials (as parameterized by $\eta$) that can be probed by the new experiment.

The increased sensitivity to chameleon models is an important feature of CHASE.  This new experiment will be sensitive to an entire class of chameleon dark energy models, as well as a wide range of power-law chameleon models.  In particular, CHASE will be able to probe the quartic chameleon, $V(\phi) = \frac{\lambda}{4!} \phi^4$.  Although this model cannot explain the observed dark energy, it has a well-understood and renormalizable potential, making it an interesting experimental target.

Quartic chameleons have been discussed extensively in the literature.~\cite{GubserKhoury,UpadhyeGubserKhoury,chamstrong2,chamstrong1}  Theories with unit self-couplings and matter couplings have been ruled out by fifth force constraints from torsion pendulum experiments.~\cite{Adelberger2007}  However, such experiments have a blind spot.  Since the source and test masses in the relevant fifth-force experiments are separated by a metal foil to keep them electrostatically isolated, a chameleon that is sufficiently massive inside metal will be screened by this foil.~\cite{chamstrong2}  Chameleons whose Compton wavelengths inside the foil are much less than the thickness of the foil will be invisible to such torson pendulum experiments.  It is precisely these chameleons which will be trapped in the CHASE chamber.  CHASE (and GammeV before it) complements laboratory searches for fifth forces.  It is also complementary to searches for a varying fine structure constant.  If we add a $\phi^4$ potential and a Yukawa matter coupling to a simple model of $\alpha$ variation such as [\refcite{Bekenstein1982}], then constraints from CHASE will be more powerful than those from laboratory or cosmological tests in a large portion of the parameter space.

\subsection{ALP Search Using the Resonant Regeneration of Photons}

An important breakthrough in increasing the sensitivity of laser experiments to ALPs was realized by [\refcite{sikivievanbibber}].  The proposed modification employs two Fabry-Perot optical cavities, one in each of the two magnetic field regions of an LSW experiment.  The first cavity serves to increase the power on the production side of the experiment.  This alone is not new as optical recycling cavities were employed in the original BFRT experiment.  The sensitivity is improved by a factor of the fourth root of the finesse of the cavity.

The second cavity, on the regeneration side, is where the breakthrough lies.  If the second optical cavity is phase-locked to the first such that it both resonates at the same frequency and with the same phase, then axions produced in the first cavity will resonantly drive the second.  This increases the sensitivity again by the fourth root of the finesse.  Thus, with two coupled Fabry-Perot cavities the sensitivity of an LSW experiment effectively improves with the square root of the finesse (it is actually the fourth root of the product of the two finesses)---a modest finesse of 100 for both cavities gives an order-of-magnitude improvement.

\begin{figure}[ht]
\centerline{\psfig{file=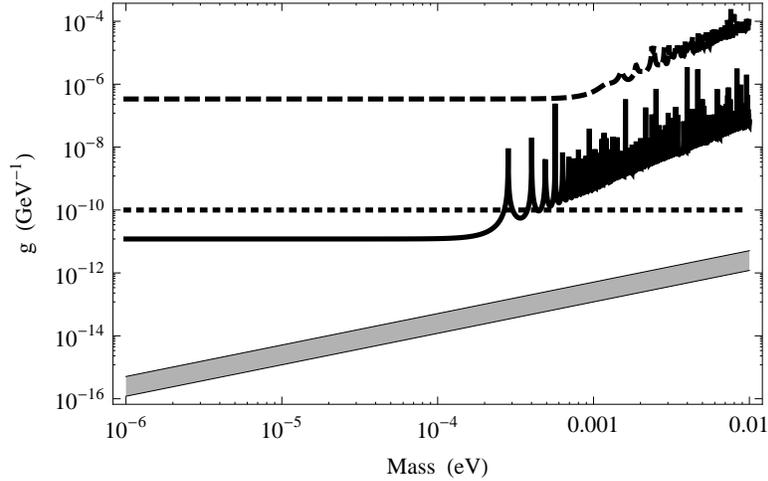,width=4in}}
\vspace*{8pt}
\caption{Anticipated sensitivity of a resonantly enhanced LSW experiment (solid line) using two strings of six, 4.2 Tesla Tevatron magnets and with cavity finesses near 100,000.  The dashed curve shows the current GammeV constraints; the dotted curve shows the limits currently set by the CAST experiment; the shaded band contains the expected QCD axion models.\protect\label{rrlimits}}
\end{figure}

An active collaboration has formed around this idea.  Using two strings of six Tevatron dipole magnets and with a finesse of a few times 10,000 will yield laboratory results comparable to those of the CAST experiment.  We believe that a larger finesse is feasible and that this technique could ultimately yield the limits shown in figure \ref{rrlimits}.~\cite{mueller2009}  While not sensitive to the QCD axion, the resonant regeneration ALP search would be sensitive to models that explain the unusual transparency of the Universe to high energy photons.~\cite{hooperserpico}  In addition, the resonant apparatus will provide nearly a five order-of-magnitude improvement in sensitivity to paraphotons.

\section{Conclusion}

The goal of the original GammeV experiment was to test the particle interpretation of the PVLAS signal.  From that kernel sprang a suite of experiments to look for low-mass particles that couple to photons.  The second installment of the suite was the original chameleon particle search.  Now, the members of the GammeV collaboration are building an improved Chameleon Afterglow Search (CHASE) and are designing an improved light-shining-through-walls search for axion-like particles---one which uses resonant cavities to increase the sensitivity to smaller photon-ALP couplings.

The GammeV ALP search excluded the particle interpretation of the PVLAS signal with high confidence, and the chameleon search was the first laboratory experiment to probe for chameleon dark-energy particles.  The improved experiments, currently under development, will increase the region of sensitivity to new ALP or chameleon particles by several orders of magnitude.  The CHASE experiment will be sensitive to chameleon dark energy models for chameleon masses beyond $\sim 1$meV as well as a wide range of power law models.  The resonant-regeneration axion experiment should make a dramatic step in sensitivity to axion-like particles, surpassing the current sensitivity of helioscope experiments in a purely laboratory setting.

Beyond seaches for ALPs and chameleons, other particles or effects may be visible to future laser experiments.  Two examples are paraphotons and mini-charged particles.~\cite{ahlers2008}  The resonant-regeneration axion experiment will also have significantly improved sensitivity to such particles.  As improvements in experimental techniques, technology, and theoretical developments materialize, laser experiments will be sensitive to an ever-increasing class of photon-coupled field theories.

\section*{Acknowledgements}

J.S. is supported by the Brinson Foundation and the Fermilab Center for Particle Astrophysics, under the U.S. Department of Energy contract DE-AC02-07CH11359.  A. U. is supported by the Kavli Institute for Cosmological Physics at the University of Chicago through grants NSF PHY-0114422 and NSF PHY-0551142 and an endowment from the Kavli Foundation and its founder Fred Kavli.



\end{document}